\documentclass[aps,pra,reprint,amsmath,amssymb,graphicx,longbibliography]{revtex4-2}

\usepackage{bm}

\usepackage[normalem]{ulem}
\usepackage[colorlinks=true,allcolors=blue,bookmarks=fals e,pdfusetitle]{hyperref}
\usepackage{url}
\usepackage{xcolor}
\usepackage{graphicx}
\usepackage{breqn}
\usepackage{braket}

\makeatletter
\let\cat@comma@active\@empty
\makeatother

\begin{document}

\title{How to measure the free energy and partition function from atom-atom correlations}

\author{Matthew L. Kerr}
\affiliation{School of Mathematics and Physics, University of Queensland, Brisbane,  Queensland 4072, Australia}
\author{Karen V. Kheruntsyan}
\affiliation{School of Mathematics and Physics, University of Queensland, Brisbane,  Queensland 4072, Australia}

\date{\today{}}

\begin{abstract}
We propose an experimental approach for determining thermodynamic properties of ultracold atomic gases with short-range interactions. As a test case, we focus on the one-dimensional (1D) Bose gas described by the integrable Lieb-Liniger model. The proposed approach relies on deducing the Helmholtz or Landau free energy directly from measurements of local atom-atom correlations by utilizing the inversion of a finite-temperature version of the Hellmann-Feynman theorem. We demonstrate this approach theoretically by deriving approximate analytic expressions for the free energies in specific asymptotic regimes of the 1D Bose gas and find excellent agreement with the exact results based on the thermodynamic Bethe ansatz available for this integrable model.
\end{abstract}

\maketitle

\section{Introduction}
Measurements of thermodynamic properties of interacting many-body systems play a critical role in characterizing and understanding the underlying physics of such systems. As an example, measurements of isothermal compressibility, $\kappa_T$, through either the measurement of atom number fluctuations \cite{Bouchoule_fluctuations_2006,Bouchoule2010:three-body,Bouchoule_dimensional_crossover,Bouchoule_sub_Poissonian,Esslinger_fluctuations_2010,Ketterle_fluctuations_2010,Chin_E0S_2011} or the density profiles in well mapped-out trapping potentials \cite{ho2010obtaining,Salomon_Thermodynamics_2010,Salomon_EoS_2010,Ueda_EoS_2010,Zwierlein:EoS,Vale_EoS_2016,1D_entropy_2017,Salces-Carcoba2018}, have been an indispensable tool in the study of strongly interacting quantum gases. This is because the isothermal compressibility is a thermodynamic quantity that can be used to deduce the equation of state (EoS) for, e.g., the mean particle number density $n=n(\mu,T)$ as a function of the chemical potential $\mu$ and the temperature $T$ of the gas. The EoS itself can then be further manipulated and used to deduce the pressure $P$ of the gas, the entropy $S$, and eventually the Helmholtz free energy $F$ or the grand potential $\Omega$ (Landau free energy), which play the central role in statistical mechanics and quantum many-body physics.

In this work, we propose an alternative experimental approach for determining thermodynamic properties of ultracold atomic gases with short-range $s$-wave scattering interactions. The approach relies on deducing the Helmholtz or Landau free energy directly from the measurements of the local (same point) atom-atom correlation function $g^{(2)}$. This is aligned more closely with the formalism of statistical mechanics, wherein one first calculates the canonical (or grand-canonical) partition function $Z$ ($\mathcal{Z}$) and the Helmholtz (Landau) free energy $F\!=\!-k_BT\ln Z$ ($\Omega\!=\!-k_BT\ln \mathcal{Z}$), and then uses these to derive the corresponding equations of state and other thermodynamic quantities, such as the entropy, pressure, or isothermal compressibility.

For simplicity and definiteness, we will illustrate this approach using the example of an ultracold one-dimensional (1D) Bose gas with short-range interactions that can be characterized by the 
$s$-wave scattering length $a$ and described by the integrable Lieb-Liniger model \cite{Lieb-Liniger-I,Lieb-Liniger-II,Olshanii1998}. We point out, however, that the approach can generally be applied to two and three-dimensional systems as well. Our proposal for determining the free energy directly from the measured local pair correlation function $g^{(2)}$ relies on the reversal of the extended version of the Hellmann-Feynman theorem for finite-temperature thermal equilibrium states (instead of the original form of the theorem \cite{hellmann1933rolle,Feynman:1939}, which was for the zero-temperature ground state). The extended form of the Hellmann-Feynman theorem was utilized in Refs.~\cite{Kheruntsyan2003,Kheruntsyan2005} for calculating the pair correlation function $g^{(2)}$ of a finite-temperature uniform 1D Bose gas from the Helmholtz free energy. The Helmholtz free energy itself was evaluated numerically using the exact Yang-Yang thermodynamic Bethe ansatz (TBA) \cite{Yang-Yang}. In the present work, we will instead assume that the pair correlation function $g^{(2)}$ can be measured experimentally, such as from photoassociation rates \cite{Weiss_g2_2005}, over a range of interaction strengths. The reversal of the Hellman-Feynman theorem then corresponds to integrating the pair correlation function over that same range of interaction strength, which in turn is equivalent to deducing the free energy of the gas. 

As a proof-of-principle illustration of the proposed approach, we will show how to derive the free energy from a known $g^{(2)}$ in four out of a total of six different asymptotic regimes of the 1D Bose gas \cite{Kheruntsyan2003,Kheruntsyan2005}. In these asymptotic regimes, the $g^{(2)}$ function can be calculated analytically using alternative approximate theoretical techniques---without resorting to a prior knowledge of the free energy and the use of the Hellmann-Feynman theorem. Such an alternative calculation of the $g^{(2)}$ function is effectively equivalent to a prior knowledge of $g^{(2)}$ from an experimental measurement, which can then be used to deduce the free energy using the same integration step, albeit now numerical.

The calculated free energy and the ensuing thermodynamic properties \cite{kerr2023analytic} have not been previously known analytically in two out of the four asymptotic regimes treated here, whereas in the two remaining regimes they reproduce the previously known results derived from the excluded volume model in the strongly-interacting regime \cite{DeRosi2022anomaly}.

We compare our approximate analytic results for the free energy with the exact numerical results obtained using the Yang-Yang TBA, and demonstrate excellent agreement. We also explain why the same calculation cannot be carried out in the remaining two (out of a total of six) asymptotic regimes. We emphasize though that while this limitation is only due to the applicability of the analytic approximation in the two regimes in question, the experimental extraction of the free energy from the measured pair correlation function is not restricted to any particular regime of the 1D Bose gas. Instead, the method only requires that the free energy is \emph{a priori} known in one of the bounds (lower or upper) of the range of the interaction strength over which the pair correlation is measured. In practice, the role of these bounds can be taken by, e.g., the ideal (noninteracting) Bose gas limit, or the Tonks-Girardeau limit of infinitely strong interactions, where the free energy is the same as that for an ideal Fermi gas by the Fermi-Bose mapping \cite{Girardeau1960,Girardeau1965,Girardeau2000,Yukalov2005,Minguzzi_Gangardt}.

\section{Preliminaries}
\subsection{The Lieb-Liniger Model}

\begin{figure*}[tbp] 
    \includegraphics[width=14.8cm]{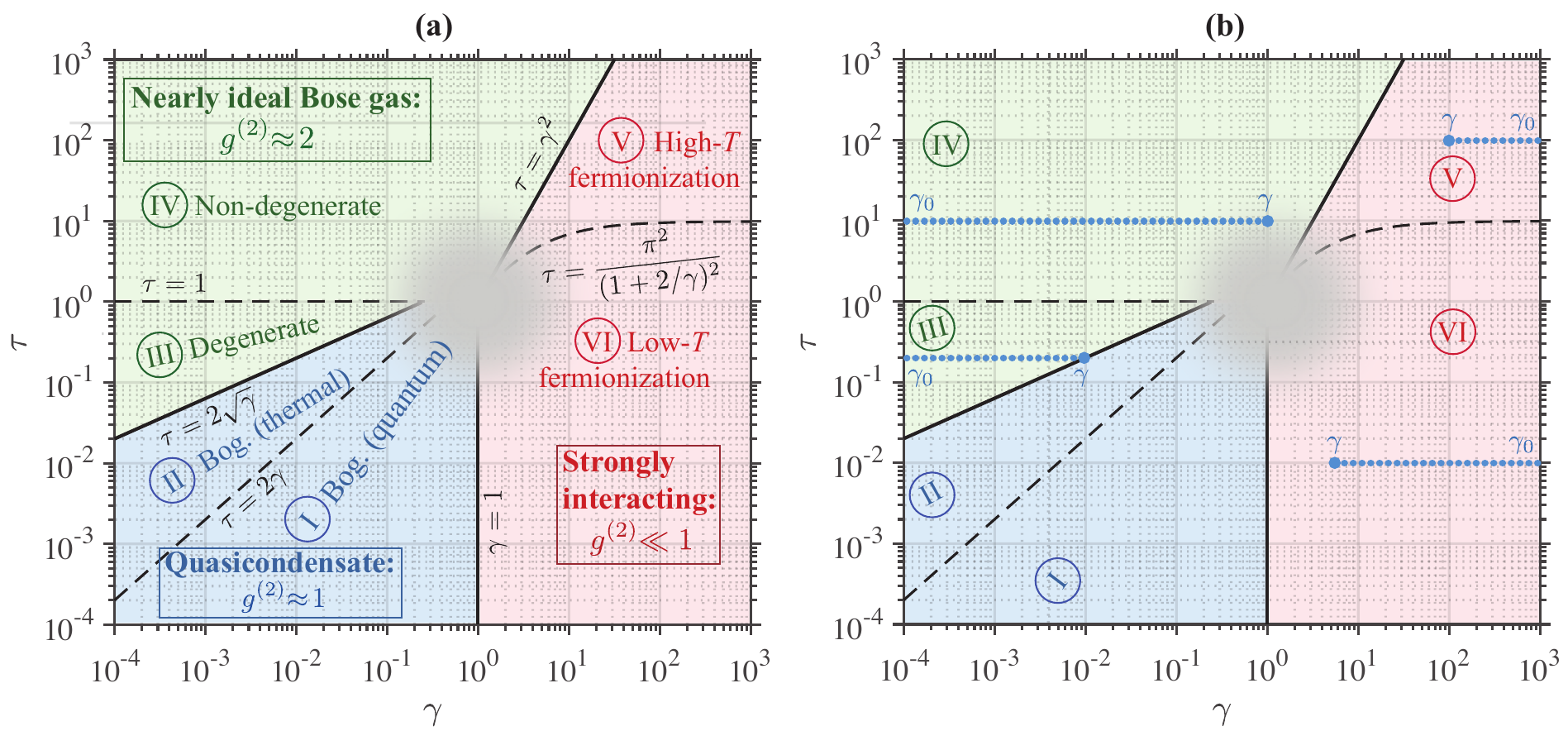}
    \caption{Crossover regimes diagram for the Lieb-Liniger gas at thermal equilibrium in the $(\gamma, \tau)$ parameter space. We distinguish six different asymptotic regimes, separated by smooth crossovers, corresponding to: the quasicondensate regime (where $g^{(2)}\simeq 1$, \cite{Kheruntsyan2003}), which can be further subdivided into the quantum (I) and thermal (II) quasicondensate regimes, respectively; the nearly ideal Bose gas regime (where $g^{(2)}\simeq 2$, \cite{Kheruntsyan2003}), which can be further subdivided into degenerate (III) and non-degenerate (IV) regimes; and the strongly interacting regime (where $g^{(2)}\ll 1$, \cite{Kheruntsyan2003}), which can be further subdivided into regions V and VI, corresponding to hiwgh-temperature and low-temperature fermionization, respectively. The blurred region in the middle of the diagram, as well as the vicinity of the regime boundaries (solid and dashed lines), is where the different analytic approximations are expected to break down.  Integration paths in Eq.~\eqref{F_integral} are shown as horizontal (blue) dotted lines, in four asymptotic regimes, in which we calculate the reduced Helmholtz free energy $\mathcal{F}(\gamma,\tau)$ and illustrate the results in Fig.~\ref{fig:F-canonical-plot} below.}
    \label{fig:regimes}
\end{figure*}

We start by considering the Lieb-Liniger Hamiltonian describing a uniform 1D gas of $N$ bosons of mass $m$ interacting  via a pair-wise $\delta$-function potential, on a line of length $L$ with periodic boundary conditions, with a linear 1D density of $n=N/L$. In second-quantized form, this is given by
\begin{align}
	\hat{H}	= & \frac{\hbar^{2}}{2m}  \int dx\, \partial_x \hat{\Psi}^{\dagger} \partial_x \hat{\Psi} + \frac{g}{2} \int dx\, \hat{\Psi}^{\dagger} \hat{\Psi}^{\dagger} \hat{\Psi} \hat{\Psi},\label{eq:H}
\end{align}
where $g$ quantifies the strength of atom-atom interactions, assumed to be repulsive ($g>0$); it can be expressed in terms of the 3D $s$-wave scattering length $a$ via $g\approx2\hbar\omega_{\perp}a$ \cite{Olshanii1998}, away from a confinement induced resonance, where $\omega_{\perp}$ is the frequency of the harmonic potential in the transverse (tightly confined) dimension.

It is convenient to define the two dimensionless quantities, 
\begin{equation}
    \gamma = \frac{mg}{\hbar^2n},\qquad \tau = \frac{2mk_BT}{\hbar^2 n^2},
\end{equation}
characterizing the interaction strength between atoms and the temperature of the system, respectively. These two parameters completely characterize the thermodynamic properties of a uniform 1D Bose gas.

\subsection{Atom-atom correlations and regimes of the uniform Lieb-Liniger gas}

In a 1D system, the normalized two-point atom-atom correlation function can be defined in terms of the bosonic field creation and annihilation operators, $\hat{\Psi}^{\dagger}(x)$ and $\hat{\Psi}(x)$, respectively, corresponding to the expectation value of a normally-ordered product of two density operators, $\hat{n}(x)=\hat{\Psi}^\dagger(x)\hat{\Psi}(x)$ and $\hat{n}(x')=\hat{\Psi}^\dagger(x')\hat{\Psi}(x')$, 
\begin{equation}
    g^{(2)}(x,x') = \frac{\langle \hat{\Psi}^\dagger(x)\hat{\Psi}^\dagger(x')\hat{\Psi}(x')\hat{\Psi}(x) \rangle}{n(x)n(x')},
\end{equation}
which is normalized to the product of mean densities $n(x)=\langle \hat{n}(x)\rangle$ and $n(x')=\langle \hat{n}(x')\rangle$ at points $x$ and $x'$. Furthermore, as we are restricting ourselves to a uniform system that is translationally invariant [with $n(x)\!=\!n(x')\!\equiv\!n$], the two-point pair correlation $g^{(2)}(x,x')$ can only depend on the relative distance $|x-x'|$, i.e., $g^{(2)}(x,x')\!=\!g^{(2)}(|x-x'|)$. The local or same-point correlation then corresponds to $x\!=\!x'$, and hence to $g^{(2)}(0)\equiv g^{(2)}$,
\begin{equation}\label{eq:g2-definition}
    g^{(2)} = \frac{\langle \hat{\Psi}^{\dagger}(x)\hat{\Psi}^{\dagger}(x) \hat{\Psi}(x)\hat{\Psi}(x)\rangle }{n^2}.
\end{equation}

As was shown in Ref.~\cite{Kheruntsyan2003} (see also \cite{Kheruntsyan2005,Sykes_2008,Deuar_Sykes_2009,kerr2023analytic} for further details), the the 1D Bose gas described by the above Lieb-Linger model can be characterized by six asymptotic regimes (separated by smooth crossovers), depending on the approximations made for deriving analytic forms of the normalized pair correlation function $g^{(2)}$, Eq.~(\ref{eq:g2-definition}). 
These asymptotic regimes, shown in Fig.~\ref{fig:regimes} in the $(\gamma, \tau)$-parameter space, can be broadly identified as the nearly ideal Bose gas regime, the quasicondensate regime, and the strongly-interacting regime.

The quasicondensate regime, corresponding to the weakly interacting 1D Bose gas ($\gamma\ll1$), can be treated using the Bogoliubov theory  \cite{mora-castin} and is characterized by suppressed density fluctuations, but fluctuating phase (which is unlike a 3D condensate with true long-range order). Depending on temperature $\tau$, this regime can be further subdivided into the quantum (I) and thermal (II) quasicondensate regimes, which are dominated by quantum and thermal fluctuations, respectively.

The nearly ideal Bose gas (IBG) regime can be treated using the perturbation theory with respect to $\gamma$ around a noninteracting Bose gas \cite{Kheruntsyan2003,Sykes_2008}; it is characterized by large fluctuations of both density and phase, and it can be further subdivided into quantum degenerate (III, $\tau\ll1$) and non-degenerate, nearly classical ideal gas (IV, $\tau\gg1$) regimes.

The strongly interacting regime $\gamma \gg 1$ can be treated using the perturbation theory with respect to $1/\gamma$ around a spin-polarized ideal (noninteracting) Fermi gas (IFG)  \cite{Cheon_1999,sen2003fermionic,Sykes_2008}, which is then mapped to the strongly interacting 1D Bose gas near the Tonks-Girardeau regime of $\gamma\to \infty$. This regime can be further subdivided into regions V and VI, corresponding to high-temperature (nondegenerate) and low-temperature (degenerate) fermionization, respectively.

In the ($\gamma,\tau$) parameter space, the six asymptotic regimes identified above can be defined via the following inequalities,
\begin{align}
  \text{I}:\,\,\, &\tau/2\ll\gamma\ll 1,\\
\text{II}:\,\,\, &2\gamma\ll \tau\ll 2\sqrt{\gamma},\\
\text{III}:\,\,\, &2\sqrt{\gamma} \ll \tau \ll 1,\\
\text{IV}:\,\,\, &\tau \gg \text{max}\{1,\gamma^2\},\\
\text{V}:\,\,\, &\pi^2/(1+2/\gamma)^2 \ll \tau \ll \gamma^2,\\
\text{VI}:\,\,\, &\tau\ll \pi^2/(1+2/\gamma)^2, \quad \gamma \gg 1.
\end{align}
Note that here we are using the updated, more accurate regime boundaries from Ref.~\cite{kerr2023analytic} (see also \cite{DeRosi2022anomaly}), for which we no longer ignore numerical factors of order one as was done in Ref.~\cite{Kheruntsyan2003}. In each of these regimes, the pair correlation function $g^{(2)}$ can be derived in closed analytic form, quoted below in Eqs.~\eqref{eq:g2_I}--\eqref{eq:g2_VI}.

\subsection{Free energy and the atom-atom correlation}
\label{g2fromF}

In the canonical formalism, the partition function $Z(T,N,L,g)$ can be written in terms of either the Helmholtz free energy $F$ or the Hamiltonian $\hat{H}$ via $Z = \exp(-F/k_BT) = \text{Tr}\exp(-\hat{H}/k_BT)$. By differentiating the Helmholtz free energy $F(T,N,L,g)=-k_BT\ln Z$ with respect to the interaction strength $g$, at constant $N$, $L$, and $T$, one finds that \cite{Kheruntsyan2003}
\begin{equation}
 \left(\frac{\partial F}{\partial g}\right)_{T,N,L} = \frac{1}{Z} \text{Tr}\left(e^{-\hat{H}/k_BT} \frac{\partial \hat{H}}{\partial g}\right) = \frac{L}{2} \braket{\hat{\Psi}^{\dagger} \hat{\Psi}^{\dagger} \hat{\Psi} \hat{\Psi}},
\end{equation}
and hence
\begin{equation}\label{eq:g2-canonical}
    g^{(2)} = \frac{2}{Ln^2} \left(\frac{\partial F}{\partial g}\right)_{T,N,L} = \frac{2m}{N\hbar^2 n^2} \left(\frac{\partial F}{\partial \gamma}\right)_{T,N,L}.
\end{equation}

Similarly, in the grand-canonical formalism, we 
start with the grand-canonical partition function $\mathcal{Z}(T,\mu,L,g)\!=\!\exp(-\Omega/k_BT)\!=\!\text{Tr}\exp[-(\hat{H}-\mu\hat{N})/k_BT]$, where $\Omega \!=\! F\!-\!\mu N \!=\! -PL$ is the Landau free energy, with $N\!=\!\langle \hat{N} \rangle$. By differentiating $\Omega(T,\mu,L,g)\!=\!-k_BT\ln \mathcal{Z}$ with respect to $g$, but now at constant $T$, $\mu$, and $L$, one finds \cite{Kheruntsyan2005}
\begin{equation}
    \left(\frac{\partial \Omega}{\partial g}\right)_{T,\mu,L} = \frac{L}{2} \braket{\hat{\Psi}^{\dagger} \hat{\Psi}^{\dagger} \hat{\Psi} \hat{\Psi}},
\end{equation}
and hence
\begin{equation}\label{eq:g2-grand_canonical}
    g^{(2)} = \frac{2}{Ln^2} \left(\frac{\partial \Omega}{\partial g}\right)_{T,\mu,L}.
 \end{equation}

Extracting the Helmholtz or Landau free energies by integrating the pair correlation function in Eq. \eqref{eq:g2-canonical} or \eqref{eq:g2-grand_canonical} (discussed below) depends on the experimental implementation. If the pair correlation $g^{(2)}$ is measured over a range of interaction strengths at constant temperature and particle number, then the appropriate approach is to adopt the canonical formalism and deduce the Helmholtz free energy first, before applying other thermodynamic relations to evaluate other thermodynamic quantities of interest (see, e.g., Ref.~\cite{kerr2023analytic}). However, if the pair correlation $g^{(2)}$ as a function of the interaction strength is measured at constant temperature and chemical potential, then it is natural to adopt the grand-canonical formalism and first deduce the Landau free energy from Eq.~\eqref{eq:g2-grand_canonical}.

\section{Free energies from the atom-atom correlation function}

The approach outlined in Sec.~\ref{g2fromF} was used for the first time in Ref. \cite{Kheruntsyan2003} for calculating the pair correlation function $g^{(2)}$ of a finite-temperature uniform 1D Bose gas. More specifically, the $g^{(2)}$ function was calculated by numerically differentiating the Helmholtz free energy, which itself was calculated exactly using the Yang-Yang TBA. In the same paper, the authors calculated the $g^{(2)}$ function in six asymptotic regimes of the 1D Bose gas using approximate analytic techniques, such as the Bogoliubov theory and the perturbation theories with respect to $\gamma$ and $1/\gamma$, around the ideal Bose gas and the ideal Fermi gas, respectively (with the latter case enabling one to treat the strongly interacting regime of $\gamma\!\gg\!1$).

The analytic calculations of the $g^{(2)}$ function in Ref. \cite{Kheruntsyan2003} did not rely on the knowledge of the Helmholtz free energy, and therefore can be used as prototypical data for deducing the corresponding free energy by inverting the Hellmann-Feynman theorem, i.e., by integrating the known $g^{(2)}$ functions. Employing the same procedure on an experimentally measured $g^{(2)}$ function constitutes the measurement approach that we are proposing in this work. We now demonstrate this approach by using the analytically calculated $g^{(2)}$ functions from Ref. \cite{Kheruntsyan2003}, given by:
\begin{align}
    \label{eq:g2_I}
    \text{I}:\,\,\, &g^{(2)} \!\simeq\! 1 - \frac2{\pi} \gamma^{1/2} +  \frac{\pi\, \tau^2}{24\, \gamma^{3/2}},   
    \,\,\,\,\,\, \left[\tau/2\ll\gamma\ll 1\right],\\
    \label{eq:g2_II}
    \text{II}:\,\,\,  &g^{(2)} \simeq 1 + \frac{\tau}{2\sqrt{\gamma}}\,\,,
    \quad \left[2\gamma\ll \tau\ll 2\sqrt{\gamma}\right],\\
    \label{eq:g2_III}
    \text{III}:\,\,\, &g^{(2)} \simeq 2 - \frac{4\gamma}{\tau^2},
     \quad \left[2\sqrt{\gamma} \ll \tau \ll 1\right], \\
    \label{eq:g2_IV}
    \text{IV}:\,\,\, &g^{(2)} \simeq 2 - \gamma\sqrt{\frac{2\pi}{\tau}},
    \,\,\,\left[\tau \gg \text{max}\{1,\gamma^2\}\right],\\ 
    \label{eq:g2_V}
     \text{V}:\,\,\, &g^{(2)} \simeq \frac{2\tau}{\gamma^2},
     \,\,\,\,\,\left[\frac{\pi^2}{(1+2/\gamma)^2} \ll \tau \ll \gamma^2\right],\\ 
    \label{eq:g2_VI}
    \text{VI}:\,\,\, &g^{(2)} \simeq \frac{4\pi^2}{3\gamma^2}\left(1 + \frac{\tau^2}{4\pi^2}\right),\,\,\left[\tau\!\ll\! \frac{\pi^2}{(1+2/\gamma)^2},\, \gamma \!\gg\! 1\right]. 
\end{align}

\begin{figure*}
\includegraphics[width=14.6cm]{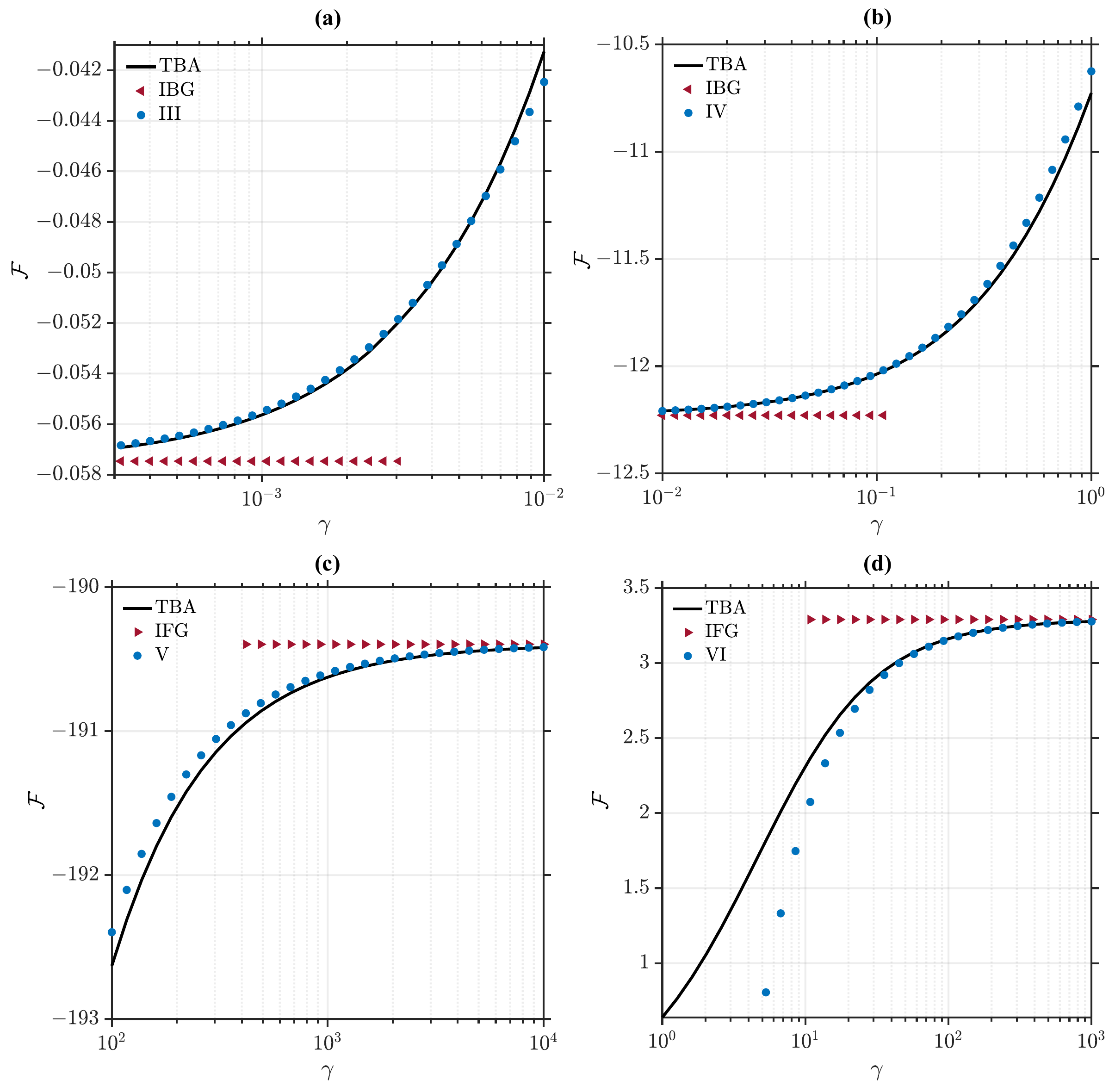}
    \caption{Helmholtz free energy per particle $F/N=\frac{\hbar^2 n^2}{2m}\mathcal{F}(\gamma,\tau)$, for: (a) $\tau=0.2$; (b) $\tau = 10$; (c) $\tau = 100$; and (d) $\tau = 0.01$. For each $\tau$, the range of $\gamma$ over which we plot our approximate analytic results is the same as the dotted horizontal lines in Fig.~\ref{fig:regimes} in regimes III--VI.} 
    \label{fig:F-canonical-plot}
\end{figure*}

\subsection{Helmholtz free energy}

Before we proceed, we note that in the canonical formalism, from dimensional considerations, the Helmholtz free energy per particle can be rewritten as
\begin{equation}
\frac{F}{N}=\frac{\hbar^2n^2}{2m}\mathcal{F}(\gamma,\tau)
\end{equation}
where $\mathcal{F}(\gamma,\tau)$ is a dimensionless function of its arguments, which is similar to Lieb and Liniger's $e(\gamma)$ corresponding to the ground state energy per particle of the 1B Bose gas, $E_0/N=\frac{\hbar^2n^2}{2m}e(\gamma)$ \cite{Lieb-Liniger-I}. Using the reduced Helmholtz free energy $\mathcal{F}(\gamma,\tau)$, Eq.~\eqref{eq:g2-canonical} can be rewritten in a dimensionless form as 
\begin{equation}\label{eq:g2-canonical-2}
    g^{(2)}(\gamma,\tau) =  \left(\frac{\partial \mathcal{F}(\gamma,\tau)}{\partial \gamma}\right)_{\tau}.
\end{equation}

In order to find the reduced Helmholtz free energy $\mathcal{F}(\gamma,\tau)$ from a known $g^{(2)}(\gamma,\tau)$, we can integrate the pair correlation function $g^{(2)}(\gamma,\tau)$ in Eq.~\eqref{eq:g2-canonical-2} with respect to the dimensionless interaction strength $\gamma$:
\begin{equation}
\mathcal{F}(\gamma, \tau) = \mathcal{F}(\gamma_0, \tau) + \int_{\gamma_0}^{\gamma} d\gamma'\, g^{(2)}(\gamma', \tau).  
\label{F_integral}
\end{equation}
Here, the reduced Helmholtz free energy at the lower bound, $\mathcal{F}(\gamma_0,\tau)$, plays the role of the integration constant and is assumed to be known for the method to work. The role of $\mathcal{F}(\gamma_0,\tau)$ can be played by, e.g., the Helmholtz free energy of the ideal Bose gas in limit of $\gamma_0=0$, i.e., $\mathcal{F}(0,\tau)\equiv\mathcal{F}_{\mathrm{IBG}}(\tau)$, or by that of an ideal Fermi gas in the opposite limit of infinitely strong interactions (the Tonks-Girardeau gas) $\gamma_0=\infty$, where $\mathcal{F}(\infty,\tau)\equiv\mathcal{F}_{\mathrm{IFG}}(\tau)$. When using the analytical results for the $g^{(2)}(\gamma,\tau)$-function in a particular regime, \eqref{eq:g2_I}--\eqref{eq:g2_VI}, we must impose an additional constraint that both integration bounds lie within the same asymptotic regime of the 1D Bose gas, where the integrand $g^{(2)}(\gamma,\tau)$ has the same functional form. This is because the analytic expressions for the pair correlation functions are valid only deep within each regime, and in particular, not at the crossover between boundaries. Consequently, the analytic approach that we are using here, whic relies solely on the knowledge of $g^{(2)}(\gamma,\tau)$ and the value of $F(\gamma_0,\tau)$ within the same analytic regime and at the same $\tau$, is only applicable in regimes III-VI, but not in regimes I and II. This restriction, however, does not apply to experimentally measured data for the correlation function $g^{(2)}(\gamma,\tau)$.

Substituting Eqs.~\eqref{eq:g2_III} and \eqref{eq:g2_IV} into Eq.~\eqref{F_integral} with $\gamma_0=0$ and carrying out the respective integrations, we obtain the Helmholtz free energy in regimes III and IV; similarly, substituting Eqs.~\eqref{eq:g2_V} and \eqref{eq:g2_VI} into Eq.~\eqref{F_integral}, but now with $\gamma_0=\infty$, and carrying out the respective integration, we obtain the Helmholtz free energy in regimes V and VI. The resulting expressions for $\mathcal{F}$ in regimes III--VI, constituting the main results of this work, are given below in Eqs.~\eqref{eq:F_III}--\eqref{eq:F_VI}. For completeness and for the benefit of the reader, we also give the results for $\mathcal{F}$ in regimes I and II, even though they have not been obtained from $g^{(2)}(\gamma,\tau)$. Instead, the Helmholtz free energy in regimes I and II can be obtained using the Bogoliubov theory by first calculating the partition function of Bogoliubov quasiparticles \cite{DeRosi2017,DeRosi2019Beyond,DeRosi2022anomaly}; within this approach, additional approximations corresponding to $\tau/2\gamma\ll1$ in regime I and $\tau/2\gamma\gg1$ in regime II allow for the derivation of simple analytic results, quoted here in Eqs.~\eqref{eq:F_I} and \eqref{eq:F_II} (derivation of these results, which extend on those reported in Refs.~\cite{DeRosi2017,DeRosi2019Beyond,DeRosi2022anomaly}, is beyond the scope of this paper, and we refer the reader to Ref.~\cite{kerr2023analytic} for details):
\begin{align}
 \text{I}&:\,\,   \mathcal{F}\simeq   \gamma - \frac{4}{3\pi}\gamma^{3/2}  - \frac{\pi}{12} \tau^2\gamma^{-1/2}  + \frac{\pi^3}{960}  \tau^4\gamma^{-5/2}\nonumber \\
 & \quad\quad\quad\quad\quad\quad\quad\quad - \frac{\pi^5}{4608} \tau^6\gamma^{-9/2},
  \label{eq:F_I}\\
 \text{II}&:\,\,  \mathcal{F} \simeq \gamma - \frac{\zeta(3/2)}{2\sqrt{\pi}} \tau^{3/2} + \tau\gamma^{1/2}  + \frac{\zeta(1/2)}{\sqrt{\pi}}\tau^{1/2}\gamma  \nonumber \\
 & \quad\quad\quad\quad\quad\quad\quad\quad\quad\quad - \frac{\zeta(-1/2)}{\sqrt{\pi}} \tau^{-1/2}\gamma^2,
\label{eq:F_II}\\
\label{eq:F_III}
\text{III}&:\,\, \mathcal{F} 
\simeq- \frac{\zeta(3/2)}{2\sqrt{\pi}} \tau^{3/2} + \frac{\tau^2}{4}
+2\gamma - \frac{2\gamma^2}{\tau^2},\\
\label{eq:F_IV}
    \text{IV}&:\,\,\mathcal{F} \simeq \tau\ln\left(\frac{2\sqrt{\pi}}{\sqrt{\tau}}\right)- \tau - \sqrt{\frac{\pi\tau}{2}} + 2\gamma - \gamma^2\sqrt{\frac{\pi}{2\tau}} ,
 \end{align}
 \begin{align}
\label{eq:F_IV}
    \text{V}:\,\,& \mathcal{F} \simeq \tau\ln\left(\frac{2\sqrt{\pi}}{\sqrt{\tau}}\right) - \tau + \sqrt{\frac{\pi \tau}{2}} -\frac{2\tau}{\gamma}, \\  \label{eq:F_VI}
    \text{VI}:\,\,&\mathcal{F} \simeq \frac{\pi^2}{3} - \frac{\tau^2}{12}  - \frac{\tau^4}{180\pi^2}  -\frac{4\pi^2}{3\gamma} - \frac{\tau^2}{3\gamma}.
\end{align}
In Eq.~\eqref{eq:F_II}, $\zeta(s)$ is the Riemann zeta function. We note that in Eq.~\eqref{eq:F_I} for regime I, the zero-temperature limit of $\tau = 0$ reproduces (as expected) the ground-state energy of the weakly interacting Bose gas, $E_0= N\frac{\hbar^2 n^2 }{2m} e(\gamma)$, with $e(\gamma)= \gamma -  \frac{4}{3\pi}\gamma^{3/2}$ \cite{Lieb-Liniger-I}.
We further note that in regimes III--VI,  the terms independent of $\gamma$ correspond to the integration constants in Eq.~\eqref{F_integral}, representing the Helmholtz free energies of the ideal Bose gas [$\mathcal{F}_{\mathrm{IBG}}(\tau)=\mathcal{F}(0,\tau)$, regimes III and IV] and the ideal Fermi gas [$\mathcal{F}_{\mathrm{IFG}}(\tau)=\mathcal{F}(\infty,\tau)$, regimes V and VI]; their derivation is outlined in the Appendix \ref{appendix:IBG}. Moreover, we point out that additional correction terms in $1/\gamma$ in the strongly interacting regimes V and VI can be derived from the excluded volume model (see \cite{DeRosi2019Beyond,DeRosi2022anomaly,kerr2023analytic} for details).

In Fig.~\ref{fig:F-canonical-plot}, we compare our approximate analytic expressions (\ref{eq:F_III})--(\ref{eq:F_VI}) with the exact Helmholtz free energy per particle calculated from the TBA. We find excellent agreement in all four regimes as $\gamma$ approaches the respective IBG or IFG boundaries.  Conversely, as $\gamma$ approaches the boundary between adjacent regimes, the analytic expressions for $g^{(2)}$ become less accurate and hence start to deviate from the TBA.

From $F(T,N,L,g)$ determined in this way, one can now deduce other thermodynamic quantities of interest, such as the pressure, entropy, chemical potential, or the isothermal compressibility in one dimension ($\kappa_T$) following the starndard prescriptions of statistical mechanics \cite{Huang_book,rev1}: $P=-(\partial F/\partial L)_{T,N,g}$, $S=-(\partial F/\partial T)_{N,L,g}$, $\mu=(\partial F/\partial N)_{T,L,g}$, or $\kappa_T=-(\partial L/\partial P)_{N,L,g}/L$, which can alternatively be determined from $\kappa^{-1}=n\,(\partial P/\partial n)_{T,N,g}$.

\subsection{Landau free energy}

From dimensional considerations in the grand-canonical formalism, the Landau free energy per particle can be rewritten as
\begin{equation}\label{eq:O_definition}
\frac{\Omega}{N}=\sqrt{\frac{\hbar^2n^2k_BT}{4m}}\mathcal{O}(\overline{\gamma},\overline{\mu}),
\end{equation}
where $\mathcal{O}$ is a dimensionless function of its arguments $\overline{\gamma}$ and $\overline{\mu}$, and we have defined a dimensionless chemical potential
\begin{equation}
\overline{\mu}=\mu/k_BT,
\end{equation}
and a new dimensionless parameter characterizing the interaction strength,
\begin{equation}
\overline{\gamma}=\sqrt{\frac{m}{\hbar^2k_BT}}\,g,
\end{equation}
using the thermal energy $k_BT$ as the energy scale.

Using the reduced Landau free energy $\mathcal{O}$, Eq.~\eqref{eq:g2-grand_canonical} can be rewritten in a dimensionless form as 
\begin{equation}\label{eq:g2-grand_canonical-2}
    g^{(2)}(\overline{\gamma},\overline{\mu}) =  \left(\frac{\partial \mathcal{O}(\overline{\gamma},\overline{\mu}))}{\partial \overline{\gamma}}\right)_{\overline{\mu}},
\end{equation}
and hence
\begin{equation}\label{eq:Omega_formula}
\mathcal{O}(\overline{\gamma}, \overline{\mu}) = \mathcal{O}(\overline{\gamma}_0, \overline{\mu}) + \int_{\overline{\gamma}_0}^{\overline{\gamma}} d\overline{\gamma}'\, g^{(2)}(\overline{\gamma}', \overline{\mu}).    
\end{equation}
Similarly to the Helmholtz free energy, the term $\mathcal{O}(\overline{\gamma}_0, \overline{\mu})$ here plays the role of the integration constant and is assumed to be known; for example, it can be taken to be the Landau free energy of the ideal Bose gas in the limit of $\overline{\gamma}_0=0$, i.e., $\mathcal{O}(\gamma_0=0, \overline{\mu})\equiv\mathcal{O}_{\mathrm{IBG}}(\overline{\mu})$, or that of the ideal Fermi gas in the Tonks-Girardeau limit, where $\mathcal{O}(\overline{\gamma}_0\to\infty, \overline{\mu})\equiv\mathcal{O}_{\mathrm{IFG}}(\overline{\mu})$.

As a simple example to illustrate this approach, we derive the reduced Landau free energy \eqref{eq:O_definition} as a function of $\overline{\gamma}$ in regime V. We first rewrite Eq.~(\ref{eq:g2_V}) for the pair correlation function $g^{(2)}$  in terms of $\overline{\gamma}$ and $\overline{\mu}$,
\begin{equation}\label{eq:GC_g2_V}
    g^{(2)}(\overline{\gamma}, \overline{\mu}) \simeq \frac{4}{\overline{\gamma}^2}.
\end{equation}
We note here that the pair correlation in this regime does not actually depend on $\overline{\mu}$ at this level of approximation, consistent with the fact that in the strictly fermionized Tonk-Girardeau limit of $\overline{\gamma}\to \infty$, the pair correlation should be exactly zero irrespective of the chemical potential and temperature due to the hard-core repulsion, which is analogous to the Pauli exclusion principle for noninteracting fermions.

By using Eq.~(\ref{eq:Omega_formula}) and identifying $\overline{\gamma}_0 = \infty$, we then find
\begin{align}\label{eq:GC_Omega_V}
    \mathcal{O}(\overline{\gamma},\overline{\mu}) &\simeq \mathcal{O}_{\text{IFG}}(\overline{\mu}) - \frac{4}{\overline{\gamma}}
    = 2\sqrt{2\pi} \frac{\text{Li}_{3/2}(-e^{\bar{\mu}})}{\left(\text{Li}_{1/2}(-e^{\bar{\mu}})\right)^2} - \frac{4}{\overline{\gamma}},
\end{align}
where $\text{Li}_{s}(z)$ is the polylogarithm function of order $s$ and argument $z$.

\begin{figure}[t]
    \centering
    \includegraphics[width=7.3cm]{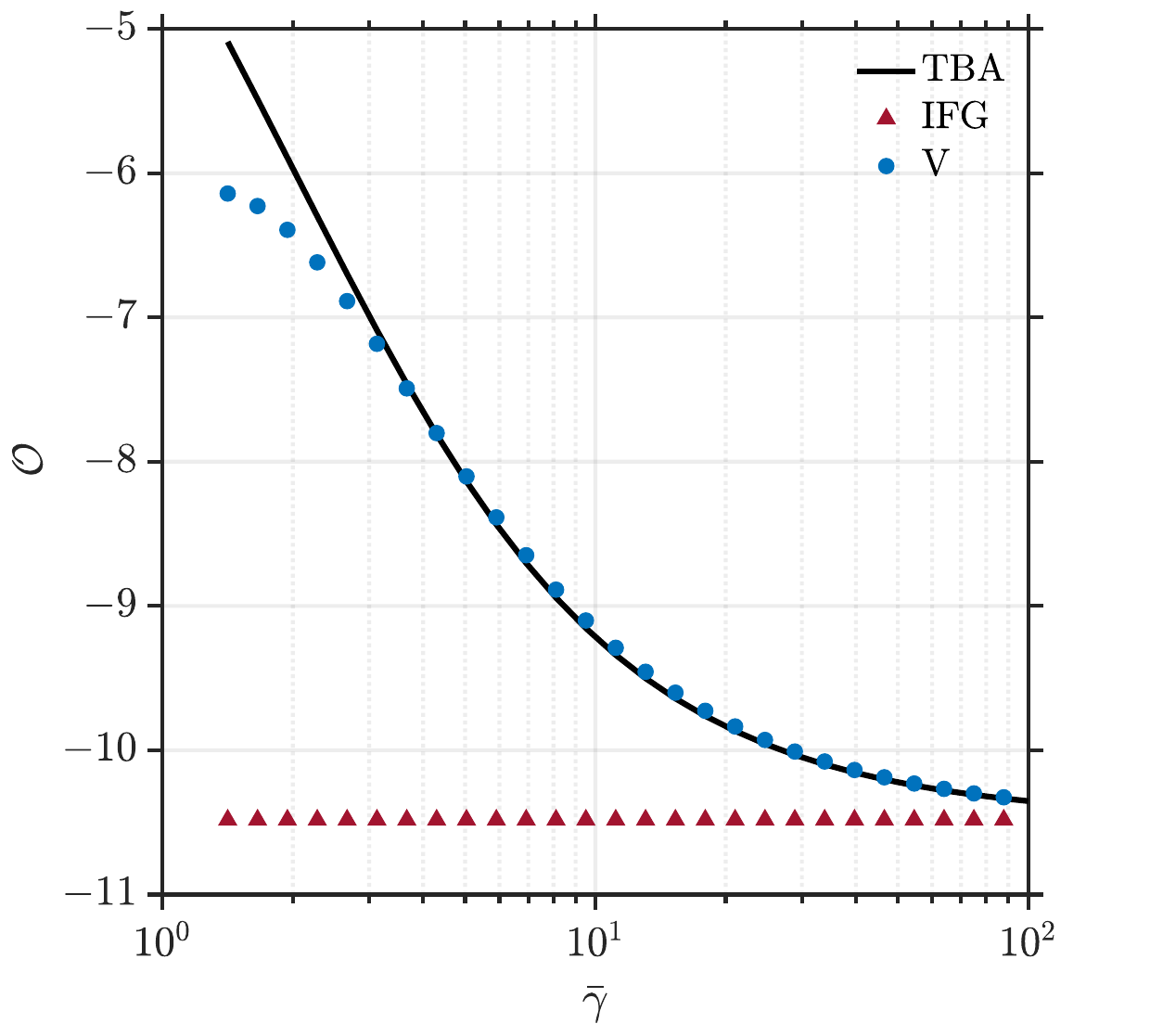}
    \caption{Landau free energy per particle, $\Omega/N=\sqrt{\hbar^2n^2k_BT/4m}\,\mathcal{O}(\overline{\gamma},\overline{\mu})$,
 as a function of $\overline{\gamma}$, for $\overline{\mu} = 0$. 
     }
    \label{fig:Omega_plot}
\end{figure}

In Fig.~(\ref{fig:Omega_plot}) we compare our analytical result with the exact predictions of the TBA by plotting the reduced Landau free energy $\mathcal{O}(\bar{\gamma},\overline{\mu})$ as a function of $\bar{\gamma}$ for a system with $\bar{\mu} = 0$, in which case
\begin{equation}\label{eq:GC_Omega_V}
    \mathcal{O}(\overline{\gamma},\overline{\mu}=0) \simeq -\frac{2\sqrt{\pi}\zeta(3/2)}{(\sqrt{2}-1)\zeta(1/2)^2} - \frac{4}{\overline{\gamma}}.
\end{equation}

We find excellent agreement between our predictions and the TBA deep within regime V. Analogous to the canonical case, however, our analytics disagree with the TBA as $\bar{\gamma}$ departs further away from the Tonks-Girardeau limit of $\overline{\gamma}\to \infty$ and towards the regime crossover boundary $\overline{\gamma} = \sqrt{2}$ (corresponding to $\tau=\gamma^2$ in Fig.~\ref{fig:regimes}\,(a)), for which the relation Eq.~(\ref{eq:GC_g2_V}) is no longer valid.

Using $\Omega(T,\mu,L,g)$ determined in this way, we can now deduce other thermodynamic quantities of interest using the standard prescriptions of statistical mechanics for grand-canonical formalism \cite{Huang_book,rev1}, such as $P=-\Omega/L$, $S=-(\partial \Omega/\partial T)_{\mu,L,g}$, or $\langle N \rangle=-(\partial \Omega/\partial \mu)_{T,L,g}$
.

\section{Summary}

We have derived approximate analytic expressions for the Helmholtz free energy of the uniform 1D Bose gas in four asymptotic regimes characterizing the system at finite temperatures and finite interaction strengths. The method relies on inverting the finite-temperature version of the Hellmann-Feynman theorem and integrating previously known analytic expressions for atom-atom pair correlation. The method can be similarly applied to derive the Landau free energy in the grand-canonical formalism. 

Our calculation can be regarded as a proof-of-principle illustration of an experimental method to deduce the free energy, and hence the ensuing thermodynamic properties of the system, from measurements of atom-atom correlations such as those utilized in Ref.~\cite{Weiss_g2_2005} using photoassociation.

Our proposed approach is complimentary to the techniques utilized in experiments of Refs.~\cite{Jin_Tan_contact_2010,Thywissen_contact_2016,Vale_contact_2019}, in which the thermodynamic properties of a strongly interacting Fermi gas were extracted from the contact parameter \cite{Minguzi_2002,Olshani_contact_2003,Tan_1,Tan_2,Braaten_Tan_2008,Zwerger_Tan_2011,Castin_contact}, which itself was measured from the high-momentum tails of the momentum distribution or the static structure factor. The complementary character of the two approaches stems from the fact that the local atom-atom correlation and contact are related by being directly proportional to each other (see, e.g., Refs. \cite{Castin_contact,Minguzzi-Laurent-Tan-contact,Bouchoule_contact_2021,DeRosi2022anomaly,De_Rosi_2023}).

\begin{acknowledgments}
The authors acknowledge stimulating discussions with Giulia De Rosi. K.\,V.\,K. acknowledges support from the Australian Research Council Discovery Project Grant No. DP190101515.
 \end{acknowledgments}

\appendix

\section{Free energy of the ideal 1D Bose and Fermi gases}
\label{appendix:IBG}

In this appendix, we outline the derivation of the Landau and Helmholtz free energies of the ideal 1D Bose gas ($\gamma= 0$) and a spin polarized ideal 1D Fermi gas, with the thermodynamic properties of the later system being identical to those of the strongly interacting 1D Bose gas in the Tonks-Girardeau limit of $\gamma\to \infty $ due to the Fermi-Bose gas mapping \cite{Girardeau1960,Girardeau1965, Girardeau2000, Castin2004, Yukalov2005, Girardeau2006}. The thermodynamic properties of the 1D IBG and IFG can be most conveniently obtained within the grand-canonical formalism of statistical mechanics \cite{Huang_book}, with the recognition that in the thermodynamic limit the expectation values of physical observables (such as the thermodynamic quantities that we are interested in this work) will be the same as in the canonical formalism.~In the grand-canonical formalism, the grand-partition function $\mathcal{Z}_{\text{IBG}}$ and the corresponding grand (or Landau) potential $\Omega_{\text{IBG}}=-k_BT\ln\mathcal{Z_{\text{IBG}}}$ for the 1D IBG or IFG are given by:
\begin{equation}
\mathcal{Z}_{\text{IBG/IFG}}=\prod_k \frac{1}{1 \mp ze^{-\epsilon(k)/k_BT}}, 
\end{equation}
and
\begin{equation}
\Omega_{\text{IBG/IFG}}=k_BT\sum_k\ln\left(1 \mp ze^{-\epsilon(k)/k_BT} \right),
\label{eq:Omega_IBG_def}
\end{equation}
where  $z = e^{\mu/\left(k_B T\right)}$ is the fugacity and $\epsilon(k)=\hbar^2k^2/2m$  is the free particle dispersion for a box of length $L$ and periodic boundary condition, with $k=(2\pi/L)j$ and $j=0,\pm 1, \pm 2,...$. Taking the thermodynamic limit, i.e., converting the discrete sum in Eq.~\eqref{eq:Omega_IBG_def} into a continuous integral, we recognize (after some algebra and integration by parts)  the resulting integral for $\Omega_{\text{IBG/IFG}}$ as the integral representation  
(also referred to as Bose function or Bose-Einstein function), 
\begin{equation}
\label{eq:polylog function}
    \text{Li}_{s}(z) = \frac{1}{\Gamma(s)}\int_{0}^{\infty}dt\frac{t^{s-1}}{e^t/z - 1},
\end{equation}
of a polylogarithm function $\text{Li}_{s}(z) = \sum_{k=1}^{\infty} z^k/k^s$ (valid for complex $s$ and $|z|<1$), with $\Gamma\left(s\right)$ being the Euler Gamma function, namely, we find that the final result for $\Omega_{\text{IBG/IFG}}$ takes the form
\begin{equation}\label{eq:Omega_IBG}
    \Omega_{\text{IBG/IFG}}  = \mp \frac{Lk_B T}{\lambda_T}\, \text{Li}_{3/2}\left(\pm e^{\mu_{\rm IBG}/k_B T}\right),
\end{equation}
where $\lambda_T=\sqrt{2\pi\hbar^2/\left(mk_BT\right)}$ is the thermal de Broglie wavelength.

The partition function $\mathcal{Z}_{\text{IBG}}=\mathcal{Z}_{\text{IBG}}(T,\mu,L,g)$  or the grand potential $\Omega_{\text{IBG}}=\Omega_{\text{IBG}}(T,\mu,L,g)$ (and similarly $\mathcal{Z}_{\text{IFG}}$ and $\Omega_{\text{IFG}}$) can be used to calculate the total thermal average number of particles in the system $\langle N \rangle$ (for a given chemical potential $\mu$) by using $\langle N \rangle= (k_BT/\mathcal{Z}_{\text{{IBG}}})(\partial \mathcal{Z}_{\text{{IBG}}}/\partial \mu)=-\partial \Omega_{\text{IBG}}/\partial \mu$; alternatively, $\langle N \rangle$ can be calculated by integrating the Bose-Einstein (Fermi-Dirac) distribution function. 
 This in turn gives  the average 1D density $n\equiv \langle n \rangle=\langle N \rangle /L$, which now takes the role of the fixed $n=N/L$ from the canonical formalism. The final result for the 1D density $n$ evaluated in this way can be written as
\begin{equation}
\label{eq:n_IBG} 
    n_{\text{IBG/IFG}}  = \pm \frac{1}{\lambda_T }\text{Li}_{1/2}\left(\pm e^{\mu/k_B T}\right).
\end{equation}

By inverting this expression for $\mu_{\text{IBG/IFG}}$, we find
\begin{equation}
\label{eq:mu_IBG_exact}
\mu_{\text{IBG/IFG}} = \frac{\hbar^2 n^2}{2m} \tau \ln \left[  \pm \text{Li}_{1/2}^{-1} \left( \pm  \frac{ 2 \sqrt{\pi}}{\sqrt{\tau}} \right)  \right], 
\end{equation}
where we have used the fact that $n\lambda_{T} = 2\sqrt{\pi/\tau}$ in dimensionless units.

We can now use the thermodynamic relation $F = \mu N+\Omega$, in addition to eliminating  $\mu$ in favor of $n$ using Eq.~\eqref{eq:mu_IBG_exact}, to find that the Helmholtz free energy per particle for the IBG and IFG is given, in terms of the 1D density and temperature, by the  expression
\begin{align}
\label{eq:F_IBG_exact}
   \frac{F_{\text{IBG/IFG}}}{N} =  & \frac{\hbar^2n^2}{2m} \left\{  \tau \ln \left[  \pm \text{Li}_{1/2}^{-1} \left(   \pm \frac{2 \sqrt{\pi}}{\sqrt{\tau}}\right) \right] \right. \nonumber \\
    & \mp  \left. \frac{\tau^{3/2}}{2 \sqrt{\pi}}  \text{Li}_{3/2} \left[ \text{Li}_{1/2}^{-1} \left( \pm \frac{2 \sqrt{\pi}}{\sqrt{\tau}}  \right)   \right]   \right\},
\end{align}
which we note is exact, valid at any temperature $\tau$.

The above expression for $F_{\text{IBG}}$ in the highly degenerate IBG regime ($\tau\ll 1$, region III)  can be further simplified using the following series  \cite{bateman1953higher} (see also \cite{Lerch}): 
\begin{equation}\label{eq:polylog_taylor_series} \text{Li}_{s}\left( e^{-\alpha}  \right) = \Gamma\left(1 - s\right) \alpha^{s - 1} + \sum_{k = 0} ^{\infty} \left(-1\right)^k \frac{\zeta\left(s-k\right)}{k!} \alpha^k,
\end{equation}
which itself is valid for $ e^{-\alpha} \approx 1$, i.e., for small and positive $\alpha \ll 1$. We let $2\sqrt{\pi/\tau}\equiv x=\text{Li}_s(e^{-\alpha})$, so that $\text{Li}^{-1}_s(x)=e^{-\alpha}$ and restrict ourselves to the first two terms in the expansion~\eqref{eq:polylog_taylor_series}, i.e. the first term and the $k\!=\!0$ term from the sum. One then obtains that $\alpha\simeq [(x-\zeta(s))/\Gamma(1-s)]^{1/(s-1)}$, and therefore $\text{Li}^{-1}_s(x)\simeq \exp{-[(x-\zeta(s))/\Gamma(1-s)]^{1/(s-1)}}$. This in turn implies
\begin{equation}
    \text{Li}_{s}^{-1}\left(\frac{2\sqrt{\pi}}{\sqrt{\tau}}\right) \simeq \exp\left[- \left(\frac{2\sqrt{\pi/\tau}-\zeta(s)}{\Gamma(1-s)}\right)^{\frac{1}{s-1}}\right].
    \label{eq:small_tau_approximation}
\end{equation}
Using Eq.~\eqref{eq:small_tau_approximation}, the reduced Helmholtz free energy $\mathcal{F}=F/N(\hbar^2n^2/2m)$ of a highly-degenerate IBG, from Eq.~\eqref{eq:F_IBG_exact}, can then be shown to be given by
\begin{align}
\label{eq:F IBG low T}
\text{III: } \mathcal{F}_{\text{IBG}} &\simeq \frac{\tau^{3/2}}{2\sqrt{\pi}} \left[ -\zeta(3/2) + \frac{\pi}{\frac{2 \sqrt{\pi}}{\sqrt{\tau}} - \zeta(1/2)}\right] +O(\tau^{5/2}) \nonumber \\
&\simeq - \frac{\zeta(3/2)}{2\sqrt{\pi}} \tau^{3/2} + \frac{\tau^2}{4} .
\end{align}
An analogous low-temperature expansion was performed within the Hartree-Fock theory in Ref.~\cite{DeRosi2022anomaly}, where the IBG case is recovered for zero coupling constant $g = 0$.

In the opposite non-degenerate limit of the IBG ($\tau\!\gg1$, region IV), one can instead use direct expansion of the Helmholtz free energy, Eq. (\ref{eq:F_IBG_exact}), in powers of the small parameter $x = 2\sqrt{\pi/\tau} \ll 1$ using the series expansion of the inverse of the polylogarithm function, $\text{Li}_{1/2}^{-1}(x)$, to the desired order. Such an expansion, in general, can be shown to be given by $\text{Li}^{-1}_s(x)= \sum_{k=1}^{\infty}a_k\,x^k$, where $a_1=1$, $a_2=-2^{-s}$, $a_3=2^{1-2s}-3^{-s}$, $a_4=56^{-s}-8^{-s}(5+2^s)$, $\cdots$  \cite{abramowitz2006handbook}. The reduced Helmholtz free energy for the non-degenerate IBG is then given by, up to the fourth-order terms (proportional to $1/\sqrt{\tau}$),
\begin{align}
\label{F_IBG_IV}
\text{IV: } \mathcal{F}_{\rm IBG} &\simeq \tau\ln\left(\frac{2\sqrt{\pi}}{\sqrt{\tau}}\right)- \tau  - \sqrt{\frac{\pi\tau}{2}} \nonumber \\
& + \pi\left(1 - \frac{4}{3\sqrt{3}}\right) + \pi^{3/2} \frac{(2\sqrt{3} - 5)}{3\sqrt{2\tau}}. 
\end{align}

Similar expansions can be performed in the nondegenerate and highly degenerate limits of the 1D IFG, except that the temperature of quantum degeneracy is now given by the Fermi temperature, $T_F=\hbar^2\pi^2n^2/(2mk_B)$. Accordingly, the nondegenerate and highly degenerate limits for the IFG correspond to $\tau\gg \pi^2$ (region V) and $\tau\ll \pi^2$  (region VI), respectively.

For region V, where $x = 2\sqrt{\pi/\tau} \ll 1$ in Eq.~\eqref{eq:F_IBG_exact}, we proceed as in region IV, i.e., using a direct series expansion of $\text{Li}_{3/2}$ in terms of its argument and the expansion of the inverse of the polylogarithm function, $-\text{Li}_{1/2}^{-1}(-x)$, which we note is positive valued. The resulting expression for the reduced free energy is nearly identical to that in Eq.~\eqref{F_IBG_IV}, except the opposite sign in front of the third term:
\begin{align}
\label{eq:high_temp_strong_F_IFG}
    \text{V: } \mathcal{F}_{\text{IFG}} &\simeq \tau\ln\left(\frac{2\sqrt{\pi}}{\sqrt{\tau}}\right) - \tau + \sqrt{\frac{\pi \tau}{2}} \nonumber \\ 
    &  + \pi\left(1-\frac{4}{3\sqrt{3}}\right) + \pi^{3/2}\frac{2\sqrt{3} - 5}{3\sqrt{2\tau}}.
    \end{align}
This result agrees with the one derived recently in Refs.~\cite{DeRosi2019Beyond, DeRosi2022anomaly} 
within the high-temperature virial expansion.

Finally, for region VI, the Helmholtz free energy of the highly-degenerate ($\tau \ll \pi^2$) IFG can be obtained using the Sommerfeld expansion \cite{ashcroft2022solid,Huang_book,DeRosi2019Beyond, DeRosi2022anomaly}, yielding
    \begin{align}
    \text{VI: } &\mathcal{F}_{\text{IFG}} \simeq \frac{\pi^2}{3} - \frac{\tau^2}{12}  - \frac{\tau^4}{180\pi^2} - \frac{7\tau^6}{1296\pi^4}.
    \label{eq:F_IFG_VI}
\end{align}


\begin{thebibliography}{59}
\bibitem [{Ler()}]{Lerch}%
  \BibitemOpen
  \href@noop {} {}\bibinfo {howpublished} {We note that the series in question
  given in Ref.~\cite{bateman1953higher} (see 1.11.8) is for the Lerch function
  $\Phi(z,s,v)=\sum_{k=0}^{\infty}z^k/(k+v)^s$. However, given that the
  polylogarithm function $\text{Li}_s(z)$ is a special case of the Lerch
  function for $v=1$, i.e. $\text{Li}_s(z)=\Phi(z,s,1)$, the series (1.11.8)
  can be adopted for the polylogarithm function, resulting in
  Eq.~\eqref{eq:polylog_taylor_series}.}\BibitemShut {Stop}%
\end{thebibliography}

%

\end{document}